\documentclass[article]{JHEP3}

\usepackage{amsmath,amssymb}
\usepackage[dvips]{graphics}
\usepackage{epsfig}

\newcommand{\be}{\begin{equation}}
\newcommand{\ee}{\end{equation}}
\newcommand{\bea}{\begin{eqnarray}}
\newcommand{\eea}{\end{eqnarray}}
\newcommand{\bean}{\begin{eqnarray*}}
\newcommand{\eean}{\end{eqnarray*}}
\newcommand{\beq}{\begin{eqnarray}}
\newcommand{\eeq}{\end{eqnarray}}
\newcommand{\nn}{\nonumber}

\relax

\def\a{\alpha'}
\def\R{{\mathcal R}}

\title{Absorption of scalars by nonextremal charged black holes in string theory}

\author{Filipe Moura
\\
Centro de Matem\'atica da Universidade do Minho, \\Escola de Ci\^encias, Campus de Gualtar, \\4710-057 Braga, Portugal\\
\\
\email{fmoura@math.uminho.pt}
}

\abstract{We analyze the low frequency absorption cross section of minimally coupled massless scalar fields by different kinds of charged static black holes in string theory, namely the D1-D5 system in $d=5$ and a four dimensional dyonic four-charged black hole. In each case we show that this cross section always has the form of some parameter of the solution divided by the black hole Hawking temperature. We also verify in each case that, despite its explicit temperature dependence, such quotient is finite in the extremal limit, giving a well defined cross section. We show that this precise explicit temperature dependence also arises in the same cross section for black holes with string $\a$ corrections: it is actually induced by them.}

\begin{document}





\section{Introduction and summary}
\indent


A classical result in black hole scattering is that in the low frequency limit, the absorption cross section of minimally coupled scalar fields by arbitrary spherically symmetric black holes is equal to the horizon area:
\be
\sigma_{\mathrm{cl}} = A_H. \label{seccaocl}
\ee
This has been shown first to some classes of four and five--dimensional extremal black holes \cite{u76,Dhar:1996vu,Das:1996wn}, and soon after to general $d-$dimensional spherically symmetric black holes \cite{dgm96}, whose metric can always be cast in the form
\be \label{schwarz}
ds^2=- f(r) dt^2+ \frac{d r^2}{g(r)}+ r^2 d\Omega^2_{d-2}.
\ee
Also in \cite{dgm96} this result has been generalized to higher spin fields, but always in Einstein gravity.

In the presence of higher derivative terms like string theoretical $\a$ corrections one must be careful, since the metric and hence the horizon area are subject to field redefinition ambiguities. In previous works \cite{Moura:2011rr,Moura:2006pz} we have studied the scattering of minimally coupled scalar fields by $d-$dimensional nonextremal spherically symmetric black holes in the presence of leading string theoretical $\a$ corrections.

Let $R_H$ be the radius of the black hole horizon: one can always choose a system of coordinates such that this quantity does not get any $\a$ corrections, i.e. $R_H\equiv\left.R_H\right|_{\a=0}.$ The dimensionless quantity $\lambda$ represents the adequate power of $\a/R_H^2$ corresponding to the leading $\a$ corrections. For a generic quantity $X$, these corrections $\delta X$ appear as a factor multiplying the same quantity computed without $\a$ corrections: $X=\left.X\right|_{\a=0} \left(1 + \lambda \ \delta X \right)$.

Let $\sigma$ be the low frequency absorption cross section and $T_H$ be the black hole temperature.\footnote{We designate the Hawking temperature by $T_H$ in order to more clearly distinguish it from other temperatures we will later consider.} In \cite{Moura:2011rr} we have concluded that one always has, to first order in $\lambda,$
\be
T_H=\left.T_H\right|_{\a=0}  \left(1 + \lambda \ \delta T_H \right), \,
\sigma=\left.\sigma\right|_{\a=0} \left(1 - \lambda \ \delta T_H \right).
\label{deltat}
\ee
This means the leading $\a$ correction to the low frequency absorption cross section is the additive inverse of the corresponding one of the temperature: $\delta \sigma = - \delta T_H.$ This $\a$ correction $\delta T_H$ is a number which depends on the concrete black hole solution one is considering, but also on the coordinate system. Indeed, (\ref{deltat}) was derived in the system of coordinates such that the horizon radius has no $\a$ corrections. In such system of coordinates, $\delta T_H$ in (\ref{deltat}) represents the only (explicit and implicit) $\a$ correction to the cross section. But the cross section should be covariant, and thus it should be expressed in terms of covariant quantities, which is certainly not the case of $\delta T_H$. From (\ref{deltat}) we see that, in order for the $(- \lambda \ \delta T_H)$ correction to $\sigma$ to appear naturally and covariantly, to first order in the perturbative parameter $\lambda,$ it must therefore come implicitly from a $1/T_H$ factor in the respective formula. This way, we are led to a covariant expression for the absorption cross section for Tangherlini--like (i.e. non--charged) black holes \cite{Moura:2011rr}:
\be
\sigma = \frac{d-3}{4 \pi T_H} \Omega_{d-2}^{\frac{1}{d-2}} A_H^{\frac{d-3}{d-2}}. \label{seccaoto}
\ee
Although it does not have an explicit dependence on $\a,$ equation (\ref{seccaoto}) encompasses the $\a$ corrections to the classical result (\ref{seccaocl}): they come implicitly, through the $\a$ corrections to $A_H$ and $T_H.$ Indeed, one can easily check that (\ref{seccaocl}) is equivalent to (\ref{seccaoto}), if this last formula is considered classically (without its implicit $\a$ corrections).

But in the presence of higher derivative terms the metric, and hence the horizon area, are subject to field redefinition ambiguities (namely field redefinitions used to cancel higher order dilaton terms that appear after the dilaton dependent conformal transformation passing from string to Einstein frame). As we have shown in \cite{Moura:2011rr}, for a Tangherlini--like $\a$-corrected black hole solution it is impossible to express the absorption cross section exclusively in terms of the mass or temperature in a way that is independent of the metric frame and of those field redefinitions: the corresponding expressions in terms of the mass or temperature in the string scheme are different than those in the Einstein scheme.\footnote{By considering further field redefinitions of order $\lambda$ to eliminate higher derivative dilaton terms that appear after changing the metric frame (e.g. from string to Einstein frame), one defines a scheme. We refer to the scheme adopted in the Einstein metric frame as Einstein scheme, and to that adopted in the string frame as the string scheme.} In the same article, we also show that the only quantity which, classically, is proportional to the black hole horizon area and for which the $\a$-corrected expressions in terms of mass and temperature are the same in both schemes is the Wald entropy. If the classical relation $\sigma= 4 G S$, obtained directly from (\ref{seccaocl}), between the low frequency absorption cross section and the Wald entropy were to be preserved by $\a$ corrections, the same would be true for the cross section: its $\a$-corrected expressions in terms of mass and temperature would be the same in both schemes. But in \cite{Moura:2011rr} we have shown that is not the case: $\a$ corrections to the entropy are manifestly different than those from the cross section. Although they are related classically, these are two intrinsically different quantities.

This is not surprising: the minimally coupled scalar field obviously does not have a stringy origin; there is no $\beta$ function from the nonlinear $\sigma-$model associated to it. Therefore, any physical property related to it, like the cross section, does not need to be invariant under metric frame transformations. On the other hand, the black hole metric by itself is a solution of string theory, and any of the black hole intrinsic properties, such as the entropy, should be invariant under such frame transformations. Nonetheless, (\ref{seccaoto}) written in that form (and not in terms of the black hole mass or charges) is valid for any metric frame.

Equation (\ref{seccaoto}) is therefore the only way to express the $\a$-corrected absorption cross section which is covariant and includes naturally the $\a$ corrections. This $1/T_H$ explicit dependence of (\ref{seccaoto}) is remarkable. As we mentioned, this formula has been derived for Tangherlini--like black holes, which do not have a finite extremal ($T_H \rightarrow 0$) limit. But, if a formula like (\ref{seccaoto}) is to be valid for more general black holes, allowing a valid finite extremal limit, as suggested in general in (\ref{deltat}), then the horizon area should have an implicit dependence on the temperature such that the absorption cross section remains finite in the limit $T_H \rightarrow 0$.

If the dependence on $\a$ of (\ref{seccaoto}) was explicit, we could split the cross section into its classical and $\a$--corrected parts, both having different functional forms. That dependence being implicit means that the functional form of the full $\a$--corrected expression is the same as the classical one. But, as we have seen, at the classical level the cross section is simply given by (\ref{seccaocl}), with just one parameter and without any explicit temperature dependence. Even if at the classical level the two expressions are equivalent, it is certainly not natural to have two different functional forms for the cross section. This means one should take (\ref{seccaoto}) already at the classical level. It would therefore be convenient to have a good physical motivation to do so, and not to take (\ref{seccaoto}) just because it is required by $\a$ corrections. That's what we will look for in this article, considering other types of classical black holes in string theory.

The two main questions we address in this article are therefore: Can a formula like (\ref{seccaoto}) for the cross section, with a $1/T_H$ explicit dependence, be obtained for more general (namely, charged) black holes? If so, is such formula well defined in the extremal limit?

The article is organized as follows. In section 2, we obtain the (simple) extension of formula (\ref{seccaoto}) to charged black holes. In section 3, we compute the low frequency limit of absorption cross section for different classes of black holes in string theory. We first consider the D1-D5 system of type IIB supergravity compactified on $S^1 \otimes T^4.$ We analyze the absorption cross section of this three-charged solution in two different regimes: with two and with one large charge. Then, we proceed analogously with a four-dimensional dyonic four-charged solution in the limit of two large charges. In all these cases we compare the obtained result for the low frequency cross section with the proposed formula of section 2, and we carefully analyze its extremal limit. We end by discussing our results.

\section{Scattering of minimally coupled scalars by spherically symmetric $\a$--corrected nonextremal black holes in $d$ dimensions}
\label{smcsss}
\indent

In this section we re-derive the result for the low frequency absorption cross section for minimally coupled massless scalar fields by general nonextremal spherically symmetric $d-$dimensional black holes with leading $\a$ corrections. The result in \cite{Moura:2011rr} was obtained for a metric of the form (\ref{schwarz}), with
\be
f(r) = c(r) f_0(r) \left(1+ \lambda f_c(r) \right), \, \, g(r)= c(r) f_0(r) \left(1+ \lambda g_c(r) \right). \label{fcgcol}
\ee
The functions $f_c(r), g_c(r)$ represent the $\a$ corrections, while the classical part was given by $c(r) f_0(r)$, with
\be
f_0(r) = 1 - \left(\frac{R_H}{r}\right)^{d-3}.
\label{tangherc} \ee
This choice implied that in Einstein gravity with $\a=0$ one had $f(r)=g(r)$ in (\ref{schwarz}); the difference between $f$ and $g$ would come only from the $\a$ corrections. This assumption is valid for $\a$-corrected Tangherlini black holes, for which one has $f(r)=g(r) \equiv f_0(r),$ i.e. $c(r) \equiv 1;$ it is also valid for $\a$-corrected Reissner-Nordstr\"om black holes, for which one has $c(r) = 1 - \left(\frac{R_Q}{r}\right)^{d-3}$, with $R_Q <R_H$.

Here we wish to re-derive the result, this time assuming the most general spherically symmetric $d-$dimensional metric, for which $f(r), g(r)$ can be two different, independent functions already for $\a=0$. The procedure is a generalization of the one taken in \cite{Moura:2011rr}; therefore we just mention the main steps, omitting the details which can be checked in that reference. The metric we take is again of the form (\ref{schwarz}), but this time with
\be
f(r) = \mathcal{F}(r) f_0(r) \left(1+ \lambda f_c(r) \right), \, \, g(r)= \mathcal{G}(r) f_0(r) \left(1+ \lambda g_c(r) \right). \label{fcgc}
\ee
Together with $f_0(r)$, $\mathcal{F}(r)$ and $\mathcal{G}(r)$ represent the classical part and are also functions of the black hole charges, while the $\a$ corrections are once more given by $f_c(r), g_c(r)$. This corresponds to the most general possible spherically symmetric metric in $d$ dimensions.
For later reference, the black hole temperature corresponding to this metric is given, to first order in $\lambda,$ by
\bea
T_H &=& \lim_{r \to R_H}{\frac{\sqrt{g}}{2\pi} \frac{d\,\sqrt{f}}{d\,r}} = \frac{\mathcal{C}(R_H)}{4\pi} \frac{d-3}{R_H} \left(1+ \lambda \ \frac{f_c(R_H) + g_c(R_H)}{2} \right), \label{temp} \\
\mathcal{C}(R_H) &:=& \sqrt{\mathcal{F}(R_H) \mathcal{G}(R_H)}.
\eea
We assume to be dealing with nonextremal black holes, but it is useful to know what is the extremal limit. Clearly, extremal black holes must verify $\mathcal{C}(R_H) \equiv 0.$

Since we are in a static, spherically symmetric background, the scalar field $\mathcal{H}$ can be expanded in terms of the spherical harmonics over the $(d-2)$ unit sphere $Y_{\ell,\varphi_1,..,\varphi_{d-3}} (\theta)$, $\ell$ being the angular quantum number associated with the polar angle $\theta$.

Even in the presence of $\a$ corrections, $\mathcal{H}(t,r)$ obeys a second order field equation of the type \cite{dgm96}
\be \label{dottigen}
\partial^2_t \mathcal{H} - F^2(r)\ \partial^2_r \mathcal{H} + P(r)\ \partial_r \mathcal{H} + Q(r)\ \mathcal{H} = 0,
\ee
$F(r), P(r), Q(r)$ being functionals of the metric (\ref{schwarz}) and its derivatives, namely of the functions $f(r), g(r).$ These functionals have $\a=0$ parts $F_{\textsf{cl}}, P_{\textsf{cl}}, Q_{\textsf{cl}},$ which by themselves have implicit $\lambda$ corrections coming from the $\lambda$ corrections of the metric. Besides those, $P, Q$ also have explicit $\lambda$--correction terms $P_{\textsf{corr}}, Q_{\textsf{corr}}$, which depend also on the $\lambda$ corrections to the Einstein equation:
\bea
F &=& F_{\textsf{cl}}, \, P = P_{\textsf{cl}} + \lambda P_{\textsf{corr}}, \, Q = Q_{\textsf{cl}} + \lambda Q_{\textsf{corr}}, \label{fpqc}\\
F_{\textsf{cl}} &=& \sqrt{fg}, \,\,
P_{\textsf{cl}} = - f \left[ (d-2) \frac{g}{r} + \frac{1}{2}
\left(f'+g'\right) \right], \nn \\
Q_{\textsf{cl}} &=& \frac{\ell \left(
\ell + d - 3 \right)}{r^2} f + \frac{(g-f)f'}{r}. \label{fpq0}
\eea
The low frequency limit of the cross section only has contributions from the mode with $\ell=0$ $\mathcal{H}_0(t,r) =: H(t,r)$ \cite{dgm96}; this is what we will take from now on.

The usual tortoise coordinate $r_*$ is defined in this case by $dr_* = \frac{dr}{F(r)}.$ We also define
\be
\Phi(t,r)= k(r) H(t,r), \,\, k(r) = \frac{1}{\sqrt{F}} \exp \left( - \int \, \frac{P}{2F^2} \, dr \right). \label{k}
\ee
Rewriting the above equation (\ref{dottigen}) in terms of $r_*$ and $\Phi(t,r),$ we obtain a wave equation with a potential $V \left[ f(r), g(r) \right]$:

\be
\frac{\partial^2 \Phi}{\partial r_*^2} - \frac{\partial^2 \Phi }{\partial t^2} = \left( Q + \frac{F'^2}{4} - \frac{F F''}{2} - \frac{P'}{2} + \frac{P^2}{4 F^2} + \frac{P F'}{F} \right) \Phi \equiv V \left[ f(r), g(r) \right] \Phi. \label{potential0}
\ee
We assume the solutions to (\ref{potential0}) to be of the form $\Phi(t,r_*) = e^{i\omega t} \Phi(r_*)$, in such a way that $$\frac{\partial^2 \Phi(t,r_*)}{\partial r_*^2} - \frac{\partial^2 \Phi(t,r_*) }{\partial t^2} =\left( \frac{d^2}{dr_*^2} +\omega^2 \right) \Phi(r_*)$$ (the same being valid for $H (t,r)$).

We are looking for a general formula for the absorption cross section, applicable to a general solution like (\ref{schwarz}). The low frequency requirement is necessary in order to use the technique of matching solutions: when the wavelength of the scattered field is very large, compared to the radius of the black hole, one can actually match solutions near the event horizon to solutions at asymptotic infinity. The idea of this technique is to separately solve the scalar field equation (\ref{dottigen}), or equivalently (\ref{potential0}), in different regions of the parameter $r,$ where in each region we take a different approximation in order to simplify the equation.

\subsection{Scattering close to the event horizon}
\label{sch}
\indent

We start by solving (\ref{potential0}) near the black hole event horizon.

We take the natural assumption for the potential $V [f(r),g(r)]$ in (\ref{potential0}): at the horizon it vanishes, and as long as $\frac{r-R_H}{R_H} \ll \left( R_H \omega \right)^2$ one will have $V [f(r),g(r)] \ll \omega^2$ and in this near--horizon region it may be neglected in (\ref{potential0}). (For a complete discussion see \cite{Moura:2011rr}.) Also $k(r)$ in (\ref{k}) is well defined at $r=R_H,$ and can be treated simply as a constant, $k(R_H),$ in a neighborhood of the horizon.

This way, we may then simply write (\ref{potential0}) as
\be
\left( \frac{d^2}{dr_*^2} +\omega^2 \right) H(r) = 0. \label{hhor}
\ee

The solutions to (\ref{hhor}) are plane waves. As we are interested in studying the absorption cross section, we shall consider the general solution for a
purely incoming plane wave:

\be \label{near}
H (r_*) = A_{\text{\tiny{near}}} e^{i \omega r_*}.
\ee

Since $f_0(R_H) \equiv 0,$ in this region the functions $f(r), g(r)$ from (\ref{fcgcol}) and $F(r)$ in (\ref{fpqc}), to first order in $\lambda,$ have the form
\bea
f(r) &\simeq& \mathcal{F}(R_H) f'_0(R_H) \left(1+ \lambda f_c(R_H) \right) \left( r-R_H \right) + {\mathcal{O}} \left( r-R_H \right)^2, \nonumber \\
g(r) &\simeq& \mathcal{G}(R_H) f'_0(R_H) \left(1+ \lambda g_c(R_H) \right) \left( r-R_H \right) + {\mathcal{O}} \left( r-R_H \right)^2, \label{fghor} \\
F(r) &\simeq& \mathcal{C}(R_H) f'_0(R_H)  \left(1+ \lambda \ \frac{f_c(R_H) + g_c(R_H)}{2} \right) \left( r-R_H \right) + {\mathcal{O}} \left( r-R_H \right)^2. \nn
\eea

This way, the tortoise coordinate, given by $r_*(r) = \int \frac{d\,r}{F(r)},$ comes as
\be
r_*(r) = \frac{R_H}{(d-3) \mathcal{C}(R_H)} \left( 1 - \lambda \frac{f_c(R_H)+g_c(R_H)}{2}\right) \log \left( \frac{r-R_H}{R_H} \right)+ {\mathcal{O}} \left( r-R_H \right),
\ee
or, in terms of the temperature (\ref{temp}), always to first order in $\lambda$,
\be
r_*(r) = \frac{1}{4 \pi T_H} \log \left( \frac{r-R_H}{R_H} \right) + {\mathcal{O}} \left( r-R_H \right). \label{expa}
\ee
Here we notice that (\ref{expa}) does not contain explicit $\lambda$ corrections: they exist, but they are implicit through the temperature dependence, by (\ref{temp}). This expression is is of course coordinate dependent; yet, this explicit dependence of the tortoise coordinate on the temperature (due to the $\a$ corrections) will be crucial for the matching process, as we will see. Replacing (\ref{expa}) in (\ref{near}), one finally obtains in this region
\be \label{close}
H (r) \simeq A_{\text{\tiny{near}}} \left( 1 + \frac{i}{4\pi} \frac{\omega}{T_H} \log \left(
\frac{r-R_H}{R_H} \right) \right) + {\mathcal{O}} \left( r-R_H \right).
\ee

\subsection{Scattering at asymptotic infinity}
\label{sai}
\indent

We consider asymptotically flat black holes which, at infinity, behave like flat Minkowski spacetime. This is equivalent to saying that, in the metric (\ref{schwarz}), functions $f(r), g(r)$ tend to the constant value 1 in the limit of very large $r,$ and the potential $V \left[ f(r), g(r) \right]$ can again be neglected. From (\ref{fcgc}), this requires
\bea
\mathcal{F}(r), \mathcal{G}(r) \underset{r \rightarrow \infty}{\longrightarrow} 1,  \label{FGfar} \\
f_c(r), g_c(r) \underset{r \rightarrow \infty}{\longrightarrow} 0.  \label{fcgcfar}
\eea

This way, in this limit (\ref{potential0}) reduces to a simple free field equation whose solutions are either incoming or outgoing plane waves in the tortoise coordinate. In the original radial coordinate, the same equation is solved in terms of Bessel functions \cite{u76,dgm96}: $H (r) = \left( r \omega \right)^{(3-d)/2} \left[ A_{\text{\tiny{asymp}}}\, J_{(d-3)/2} \left( r\omega \right) + B_{\text{\tiny{asymp}}}\, N_{(d-3)/2} \left( r\omega \right) \right].$ At low frequencies, with $r\omega \ll 1$, such solution becomes
\be \label{far}
H (r) \simeq A_{\text{\tiny{asymp}}}\ \frac{1}{2^{\frac{d-3}{2}} \Gamma \left( \frac{d-1}{2} \right)} + B_{\text{\tiny{asymp}}}\ \frac{2^{\frac{d-3}{2}} \Gamma \left( \frac{d-3}{2} \right)}{\pi \left( r\omega \right)^{d-3}} + {\mathcal{O}} \left( r\omega \right).
\ee

\subsection{Scattering in the intermediate region}
\label{sir}
\indent

We now consider the intermediate region: far from the horizon, but not asymptotic infinity. In this region the magnitude of the potential may be large.

We are working perturbatively in $\lambda.$ We then define the expansions
$H (r) = H_0 (r) + \lambda H_1 (r), k (r) = k_0 (r) + \lambda k_1 (r)$ and similarly $F = F_0 + \lambda F_1, P = P_0 + \lambda P_1, Q = Q_0 + \lambda Q_1.$ $F_0, P_0, Q_0$ represent the $\lambda=0$ parts of $F_{\textsf{cl}}, P_{\textsf{cl}}, Q_{\textsf{cl}}$ given in (\ref{fpq0}), while $F_1, P_1, Q_1$ represent the order $\lambda$ parts of $F, P, Q$: those from the order $\lambda$ parts of $F_{\textsf{cl}}, P_{\textsf{cl}}, Q_{\textsf{cl}}$ and those directly from $P_{\textsf{corr}}, Q_{\textsf{corr}}$ given in (\ref{fpqc}).

Having this expansion in mind, we expand every term of (\ref{dottigen}), assuming the condition $\omega^2 \ll V [f(r),g(r)].$ To zero order in $\lambda$ we obtain
\be \label{h0mod}
H_0'' - \frac{P_0}{F_0^2} H_0' - \frac{Q_0}{F_0^2} H_0 =0.
\ee
This is a second order linear ordinary differential equation, whose most general solution is
\be \label{h0sol}
H_0 (r) = A_{\text{\tiny{inter}}}^0 + B_{\text{\tiny{inter}}}^0 \int\frac{d\,r}{r^{d-2} F_0(r)}.
\ee
The terms of first order in $\lambda$ are given by
\be \label{h1}
H_1'' - \frac{P_0}{F_0^2} H_1' - \frac{Q_0}{F_0^2} H_1= R(r), \,\, R(r)=-\left(\frac{F_1}{F_0}\right)^2 H_0'' + \frac{P_1}{F_0^2} H_0' + \frac{Q_1}{F_0^2} H_0
\ee
This is a second order linear nonhomogeneous differential equation for $H_1.$ The homogeneous part is exactly the same as the differential equation (\ref{h0mod}) for $H_0,$ with general solution (\ref{h0sol}), obviously replacing $H_0 (r), A_{\text{\tiny{inter}}}^0, B_{\text{\tiny{inter}}}^0 $ by $H_1 (r), A_{\text{\tiny{inter}}}^1, B_{\text{\tiny{inter}}}^1 .$
To obtain the most general solution to (\ref{h1}) one just needs to add to a particular solution $H_1^{\text{\tiny{part}}} (r),$ obtained by the method of variation of constants, the most general solution (\ref{h0sol}) to the homogeneous equation (\ref{h0mod}), including the contributions $H_0, H_1$ as $H=H_0 + \lambda H_1.$ Defining
\bea
h_2(x) &=& \int\frac{d\,x}{x^{d-2} F_0(x)}, \label{h2} \\
v_1(r) &=& - \int R(r) r^{d-2} F_0(r) h_2 (r) \, d\,r, \,\, v_2(r) = \int R(r) r^{d-2} F_0(r) \, d\,r, \label{v1v2}
\eea
the solution for $H(r)$ is given by
\bea
H (r) &=& A_{\text{\tiny{inter}}} + B_{\text{\tiny{inter}}} \int\frac{d\,r}{r^{d-2} F_0(r)} + \lambda H_1^{\text{\tiny{part}}} (r) \nonumber \\
&=& \left(A_{\text{\tiny{inter}}}+ \lambda v_1(r) \right) + \left(B_{\text{\tiny{inter}}} + \lambda v_2(r)\right) \int\frac{d\,r}{r^{d-2} F_0(r)}. \label{h1mod}
\eea
Asymptotically the ``varying constants'' $v_1(r), v_2(r)$ in (\ref{v1v2}) vanish since, from (\ref{fcgcfar}), all $\lambda$ corrections vanish at infinity, including $R(r)$. Close to the black hole horizon, from (\ref{v1v2}) we can see that the contributions of $v_1(r), v_2(r)$ are subleading (for a detailed analysis see \cite{Moura:2011rr}). This means that close to these regions, we can neglect $H_1^{\text{\tiny{part}}} (r)$ and simply consider the solution to the homogeneous equation $H_0(r).$ This will be a key feature for the matching process.

\subsection{Calculation of the absorption cross section}
\label{cacs}
\indent

We are now ready to start the matching process, using $f_0$ given by (\ref{tangherc}).

If we evaluate (\ref{h1mod}) near the horizon, we obtain
\be
H (r) \simeq A_{\text{\tiny{inter}}} + \frac{B_{\text{\tiny{inter}}}}{(d-3) R_H^{d-3} \mathcal{C}\left(R_H\right)} \log \left( \frac{r-R_H}{R_H} \right) + {\mathcal{O}} \left(\frac{r-R_H}{R_H}\right).
\ee

Matching the coefficients above to the ones in (\ref{close}) immediately yields
\bea
A_{\text{\tiny{near}}} &=& A_{\text{\tiny{inter}}}, \nonumber \\
B_{\text{\tiny{inter}}} &=& \frac{i}{4\pi} \frac{\omega}{T_H} (d-3) \mathcal{C}(R_H) R_H^{d-3} A_{\text{\tiny{near}}}.
\eea

From condition (\ref{FGfar}) in section \ref{sai} we have, at asymptotic infinity, to leading order, $h_2(r) =\int\frac{d\,r}{r^{d-2} F_0(r)} \simeq
-\frac{1}{d-3} \frac{1}{r^{d-3}} + \cdots,$ and, therefore, evaluating (\ref{h1mod}) again asymptotically,
\be
H (r) \simeq A_{\text{\tiny{inter}}} - \frac{B_{\text{\tiny{inter}}}}{d-3} \frac{1}{r^{d-3}} + \cdots.
\ee
\noindent
In this region one may match the coefficients above to the ones in (\ref{far}), yielding

\bea
A_{\text{\tiny{asymp}}} &=& 2^{\frac{d-3}{2}} \Gamma \left( \frac{d-1}{2} \right) A_{\text{\tiny{inter}}} = 2^{\frac{d-3}{2}} \Gamma \left( \frac{d-1}{2} \right) A_{\text{\tiny{near}}}, \label{match} \\
B_{\text{\tiny{asymp}}} &=& - \frac{\pi \omega^{d-3}}{2^{\frac{d-3}{2}} (d-3) \Gamma \left( \frac{d-3}{2} \right)} B_{\text{\tiny{inter}}} = - \frac{i \omega ^{d-2} R_H^{d-3}}{2^{\frac{d+3}{2}} \Gamma \left( \frac{d-1}{2} \right)} (d-3) \frac{\mathcal{C}(R_H)}{T_H} A_{\text{\tiny{near}}}. \nonumber
\eea
The incoming flux per unit area is computed near the black hole event horizon where, from (\ref{near}),
\be \label{jnear}
J_{\text{\tiny{in}}} = \frac{1}{2i} \left( H^\dagger (r_*) \frac{d H}{d r_*} - H (r_*) \frac{d H^\dagger}{d r_*} \right) = \omega \left| A_{\text{\tiny{near}}} \right|^2.
\ee
The outgoing flux per unit area is computed at asymptotic infinity, where $r_*$ and $r$ coincide and, from (\ref{far}),
\be
J_{\text{\tiny{out}}} = \frac{1}{2i} \left( H^\dagger (r) \frac{d H}{dr} - H (r)
\frac{d H^\dagger}{dr} \right) =\frac{2}{\pi} r^{2-d} \omega^{3-d} \left| A_{\text{\tiny{asymp}}}
B_{\text{\tiny{asymp}}} \right|.
\ee
The absorption cross section is given by $$\sigma = \frac{\int r^{d-2} J_{\text{\tiny{out}}}d \Omega_{d-2}}{J_{\text{\tiny{in}}}}= \frac{2}{\pi}  \omega^{2-d} \frac{\left| A_{\text{\tiny{asymp}}}
B_{\text{\tiny{asymp}}} \right|}{\left| A_{\text{\tiny{near}}} \right|^2} \Omega_{d-2}.$$ Replacing the results from (\ref{match}), and after some simplifications, the final result is
\be
\sigma = (d-3)\frac{\mathcal{C}(R_H)}{4 \pi T_H} \Omega_{d-2}^{\frac{1}{d-2}} A_H^{\frac{d-3}{d-2}}, \label{seccaoc}
\ee
where $A_H=R_H^{d-2} \Omega_{d-2}$ is the horizon area with respect to the metric induced by (\ref{schwarz}).

We see that in the formula for the cross section, for charged black holes, the $1/T_H$ factor present in (\ref{seccaoto}) gets replaced by a $\mathcal{C}(R_H)/T_H$ factor. In the extremal limit, both the numerator and the denominator of this factor go to 0, but from (\ref{temp}) we learn that $\frac{\mathcal{C}(R_H)}{T_H} =\frac{4 \pi R_H}{(d-3)\left(1+ \lambda \ \frac{f_c(R_H) + g_c(R_H)}{2} \right)},$ a finite result. This way, because of the $\mathcal{C}(R_H)$ factor in the numerator, the low frequency absorption cross section remains finite in the extremal limit.

For nonextremal black holes, one can always re-scale the time as $d\tilde{t}=\mathcal{C}(R_H) \, dt$ (and equivalently, after euclideanization, the time periodicity, given by $1/T_H$), as explained in \cite{Moura:2011rr}. This means that time and temperature can always be chosen in order to set $\mathcal{C}(R_H) \equiv 1$, and therefore (\ref{seccaoc}) reduces to (\ref{seccaoto}). Apart from this detail, the result for the cross section for arbitrary nonextremal charged black holes is analogous to the one for noncharged black holes, given by (\ref{seccaoto}), and the discussion on the temperature dependence following it is also valid.

\section{The absorption cross section for charged black holes in string theory}
\indent

In this section we take examples of charged black holes in the context of string theory: first, we consider three-charged five dimensional black holes in different regimes; next, we consider analogous four dimensional black holes with four charges. In each case, we compute the corresponding low frequency absorption cross section, naturally obtaining a $1/T_H$ explicit dependence. We analyze each expression in the extremal limit. Finally we compare and discuss our results.

\subsection{Three-charged five-dimensional black holes}
\indent

First we review a class of charged black hole solutions of type IIB supergravity, the D1-D5 system. Setting to zero all the fermions as usual, but also the NS-NS two-form, the R-R four-form with self-dual field strength and the R-R scalar, we are left with the following low energy effective action (in the string scheme):
\be
\frac{1}{16 \pi G_{10}}
\int \sqrt{- g} \left[ e^{-2\phi} \left( \R+4(\nabla \phi )^2 \right)
-\frac{1}{12} \left\vert H_{3} \right\vert^2 \right] d^{10}x. \label{mea}
\ee
Here $16 \pi G_{10} = (2 \pi)^7 \ell_{s}^{8} g_s^2$, $g_s$ being the string coupling constant and $\ell_{s}$ the string scale. We take the effective action (\ref{mea}) compactified on a circle $S^1$ of length $2\pi R$ and a torus $T^4$ of four-volume $(2\pi)^4 V.$

The solutions for the metric and dilaton are given by \cite{Horowitz:1996ay}
\bea
ds_{10}^2 &=& \frac{1}{\sqrt{h_{1}(r) h_{5}(r)}} \left( -f(r) {dt^\prime}^2 + {dx_5^{\prime}}^2 \right)
+ \sqrt{\frac{h_{1}(r)}{h_{5}(r)}} dx_i dx^i \nn \\
&+&\sqrt{h_{1}(r) h_{5}(r)} \left( \frac{dr^2}{f(r)} + r^2 d \Omega_{3}^2 \right), \\
e^{2\phi } &=& \frac{h_{5}(r)}{h_{1}(r)},
\eea
where we define the thermal and harmonic functions
\be
f(r) = 1-\frac{r_0^2}{r^2}, \, h_{1,5}(r) = 1 + \frac{r_{1,5}^2}{r^2}  \label{fh}
\ee
and the boosted coordinates
\be
\left( \begin{array}{c}
t^\prime\\ x^{5\prime} \end{array} \right) =
\left( \begin{array}{cc} \cosh \varsigma & \sinh \varsigma \\
\sinh \varsigma & \cosh \varsigma  \end{array} \right)
\left( \begin{array}{c} t\\ x^5 \end{array} \right). \label{boost}
\ee
$x^5$ is periodically identified with period $2\pi R$, and $x^i$, $i = 6,...,9$, are each identified with period $2\pi V^{1/4}$.

The R-R 2-form field strength is given by
\be
H_{(3)} = 2 r_5^2 \epsilon_3 + 2 e^{-2\phi} r_1^2 \star_6 \epsilon_3.
\ee
Here $\epsilon_3=\frac{1}{8} d\theta \wedge \sin \theta d \phi \wedge d \psi$ is the volume element on the unit three-sphere, and $\star_6$ is the Hodge dual in the six dimensions $x^0,..,x^5$.

One can define $r_n$ in terms of $r_0$ and the parameter $\varsigma$ in (\ref{boost}) by
\be
r_n \equiv r_0 \sinh\varsigma. \label{rn}
\ee
One can introduce similar parameters $\alpha, \gamma$ in order to similarly relate $r_1, r_5$ to $r_0$:
\bea
r_1 &\equiv& r_0 \sinh\alpha, \label{r1} \\ r_5 &\equiv& r_0 \sinh\gamma. \label{r5}
\eea
Besides the compactification moduli $R$ and $V$, this solution has four independent parameters
$r_0, \, r_1, \, r_5, \, r_n,$ or equivalently $r_0, \, \alpha, \, \gamma, \, \varsigma$, in terms of which one may write the black hole mass and three $U(1)$ charges. Two of such charges are the quantized fluxes of the $H_{(3)}$-form and its dual:
\be
Q_1 \equiv \frac{1}{(2 \pi)^2 \, g_s} \int_{S^3} e^{2\phi} \star_6 H_{(3)}, \, \,
Q_5 \equiv \frac{1}{(2 \pi)^2 \, g_s} \int_{S^3} H_{(3)},
\ee
The third charge in the model is the quantized momentum around $S^1$ (in this case along the compact $x^5$ direction): $n = R P.$ All charges are normalized to be integers and taken to be positive. In terms of the parameters of the solution, the charges are given by
\bea
Q_1 &=& \frac{V}{2 g_s} r_0^2 \sinh 2\alpha, \label{q1} \\
Q_5 &=& \frac{1}{2 g_s} r_0^2 \sinh 2\gamma, \label{q5} \\
n &=& \frac{R^2 V}{2 g_s^2} r_0^2 \sinh 2\varsigma. \label{n}
\eea

The D-brane description of this black hole \cite{Callan:1996dv} involves a bound state of $Q_1$ fundamental strings wrapping $S^1$ and $Q_5$ NS5-branes wrapping $T^4 \otimes S^1$. Excitations of this bound state are approximately described by transverse oscillations, within the NS5-brane, of a single effective string wrapped $Q_1Q_5$ times around the $S^1$. These oscillations carry the momentum $n$ and are described by a gas of left and right movers on the string.

After reduction to five dimensions, the corresponding metric is given by
\bea
ds_{5}^2 &=& -h^{-2/3}(r) f(r) d t^2 + h^{1/3}(r) \left( \frac{dr^2}{f(r)} + r^2 d \Omega_{3}^2 \right), \label{g5} \\
h(r) &=& h_1(r) h_5(r) h_n(r), \, h_n(r) = 1 + \frac{r_n^2}{r^2},
\eea
with $f(r)$ and $h_{1,5}(r)$ defined as in (\ref{fh}). The metric (\ref{g5}) is invariant under permutations of the parameters $\alpha, \gamma, \varsigma.$ This is a consequence of $U-$duality, as it was used for the obtention of this solution \cite{Horowitz:1996ay}.

We will now assume that in this background massless, neutral, minimally coupled scalar fields are absorbed. We will examine the corresponding low energy absorption cross section in two different regimes, concerning the magnitudes of the charges.
\subsubsection{Black holes with two large charges}
\label{2lc}
\indent

We now assume the following geometrical condition on the solution we are working with:
\be
r_0 , r_n \ll r_1 , r_5. \label{dgr}
\ee
This condition corresponds to having large values of the parameters $\alpha, \gamma$ in (\ref{r1}) and (\ref{r5}). In this limit, the relations (\ref{q1}) and (\ref{q5}) between the charges $Q_1, \, Q_5$ are replaced by
\be \label{r1r5}
r_{1}^2 =  g_s \frac{Q_1}{V}, \, \, r_{5}^2 = g_s Q_5.
\ee
Equation (\ref{n}) for the quantized momentum around $S^1$ remains valid. The physical meaning of condition (\ref{dgr}), in the black hole description, is that $Q_1$ and $Q_5$ are large compared to $n.$

In the string/brane description, condition (\ref{dgr}) defines the dilute gas limit. In this regime, interactions between the left and right moving oscillations we referred to can be neglected; these oscillations are governed by effective left and right moving temperatures given by
\be
T_{L} = \frac{1}{2 \pi} \frac{r_0 e^\varsigma}{r_1 r_5}, \, T_{R} = \frac{1}{2 \pi} \frac{r_0 e^{-\varsigma}}{r_1 r_5}.  \label{tltr}
\ee
They are related to the overall Hawking temperature by
\be
T_{H}^{-1} = \frac{1}{2}(T_{L}^{-1}+T_{R}^{-1}), \label{thtltr}
\ee
from where we write
\be
T_{H} = \frac{1}{2 \pi} \frac{r_0 }{r_1 r_5 \cosh \varsigma}. \label{th}
\ee

The relation (\ref{th}) can be inverted, in order to express the parameter $\varsigma$ in (\ref{boost}) as a function of $r_0, \, r_1, \, r_5$ and $T_H,$ obtaining
\be
e^\varsigma= \frac{r_0+\sqrt{r_0^2 - 4 \pi^2 r_1^2 r_5^2 T_H^2}}{2 \pi r_1 r_5 T_H}. \label{sigma}
\ee
Replacing (\ref{sigma}) in (\ref{n}), one obtains an equation expressing $n$ as a function of $r_0, \, r_1, \, r_5, \, T_H, \, R$ and $V.$ This equation may be explicitly solved for $r_0$: defining
\bea
x=\pi r_1 r_5 T_H, y= R^2 V, z= g_s^2 n, \nonumber \\
F(x,y,z) \equiv \sqrt[3]{9 x^8 y^4 z^2+\sqrt{3} \sqrt{27 x^{16} y^8 z^4+16 x^{12} y^6 z^6}}, \nonumber \\
G(x,y,z) \equiv x^4 + \frac{2\ 2^{2/3} x^4 z^2}{\sqrt[3]{3} F(x,y,z)}-\frac{\sqrt[3]{2} F(x,y,z)}{3^{2/3} y^2},
\eea
we have
\be
r_0^2= x^2+\sqrt{G(x,y,z)}
+\sqrt{\frac{2 x^6}{\sqrt{G(x,y,z)}}+3 x^4 - G(x,y,z)}. \label{r0}
\ee
This way we can express $r_0$ in terms of $n, \, r_1, \, r_5, \, T_H, \, R$ and $V,$ and independently of $\varsigma.$ This expression can be expanded in power series of the temperature:
\be
r_0^2=2\frac{g_s \pi r_1 r_5}{R} \sqrt{\frac{n}{V}} T_H +\pi^2 r_1^2 r_5^2 T_H^2 +\frac{3 \pi^3 r_1^3 r_5^3 R}{4 g_s} \sqrt{\frac{V}{n}} T_H^3 + \frac{\pi^4 r_1^4 r_5^4 R^2 V}{2 g_s^2 n} T_H^4 + {\mathcal{O}} \left(T_H^5\right). \label{r0t}
\ee
The expansion (\ref{r0t}) shows that $r_0^2$ is a smooth function with a well defined extremal ($T_H \rightarrow 0$) limit, and that limit is 0.

Replacing the solution (\ref{r0}) for $r_0^2$ in (\ref{sigma}), we see that we can also express $e^\varsigma$ in terms of $r_1, \, r_5, \, T_H, \, R$ and $V.$ In order to take the extremal limit, we should rather replace the expansion (\ref{r0t}) in (\ref{sigma}), to conclude that $\lim_{T_H \rightarrow 0} e^\varsigma= +\infty.$ Both limits of $r_0$ and of $\varsigma$ are consistent with the expression for $T_H$ obtained in (\ref{th}). But, in order to have a finite value for $n$ in the extremal limit, from (\ref{n}) one needs a finite value of $r_n$ given by (\ref{rn}). The complete geometrical definition of the extremal limit is therefore $r_0 \rightarrow 0, \, \varsigma \rightarrow +\infty,$ but in such a way that $r_n$ remains finite. In terms of the temperatures, from (\ref{tltr}) the extremal limit is defined as $T_R, T_H \rightarrow 0$ with $T_L$ remaining finite.

Because of the dilute gas assumption (\ref{dgr}), the full (frequency-dependent) absorption cross section for this solution could be computed, the result being \cite{Maldacena:1996ix}
\be
\sigma_{abs}(\omega) =
\pi^3 r_1^2 r_5^2 \omega \frac{e^{\frac{\omega}{T_H }} -1}{ \left(e^{\frac{\omega}{2T_L }} -1\right)\left(e^{\frac{\omega}{2T_R }} -1\right)}.
\label{seccaow}
\ee
We are interested in the low frequency limit of the absorption cross section, which can be obtained from the $\omega \rightarrow 0$ limit of (\ref{seccaow}).
Using the relation (\ref{thtltr}), we can write such limit in the form
\be
\sigma= 2 \pi^3 r_1^2 r_5^2 (T_L + T_R). \label{seccaowl}
\ee
This same limit can be written, using the relations (\ref{tltr}) and (\ref{th}), in the equivalent form
\be
\sigma = \frac{\pi r_0^2}{T_H}. \label{seccaowl2}
\ee
We see that the low frequency limit of the absorption cross section is of the general form for charged solutions, obtained in (\ref{seccaoc}): the quotient of a term that goes to 0 in the extremal limit ($\pi r_0^2,$ as we have just seen), and the black hole temperature. Considering that low frequency cross section written in the form (\ref{seccaowl}), it is easy from our previous discussion to conclude that also for this solution it has a smooth extremal limit. That conclusion could also have been achieved by replacing the expansion (\ref{r0t}) in (\ref{seccaowl2}).

\subsubsection{Black holes with one large charge}
\label{1lc}
\indent

We now assume a different range of parameters for the solution we are working with, defined by
\be
r_0 , r_1, r_n \ll r_5. \label{dgr2}
\ee
This condition corresponds to having a large value of the parameter $\gamma$ in (\ref{r5}) (but not of $\alpha$ in (\ref{r1}) nor of $\varsigma$ in (\ref{rn})). Physically, condition (\ref{dgr2}) means that $Q_5$ is large compared to the other charges $Q_1$ and $n.$ The temperatures of the left and right moving oscillations and the black hole Hawking temperature, verifying condition (\ref{thtltr}), are given respectively by
\bea
T_L &=& \frac{1}{2\pi r_5 \cosh(\alpha - \varsigma)}, \, T_R= \frac{1}{2\pi r_5 \cosh(\alpha+ \varsigma)}, \label{tltr2} \\
T_H &=& \frac{1}{2\pi r_5 \cosh \alpha \cosh \varsigma}. \label{th2}
\eea

In this range it is also possible to compute the full (frequency-dependent) absorption cross section, the result being \cite{Klebanov:1997cx}
\be
\sigma_{abs}(\omega)= \pi^3 r_5^2 r_0^2 (1+ \sinh^2 \alpha +\sinh^2 \varsigma) \omega \frac{e^{\frac{\omega}{T_H }} -1}{ \left(e^{\frac{\omega}{2T_L }} -1\right)\left(e^{\frac{\omega}{2T_R }} -1\right)}. \label{seccaow2}
\ee
This formula has a structure which is similar to the one of (\ref{seccaow}). Indeed, using (\ref{tltr2}) one can see that its low frequency limit is the same as the one of (\ref{seccaow}), given by (\ref{seccaowl2}): $\sigma = \frac{\pi r_0^2}{T_H}.$

In order to study the extremal limit, first we invert (\ref{q1}) (respectively (\ref{n})) in order to express $e^\alpha$ (respectively $e^\varsigma$) in terms of $Q_1$ (respectively $n$), $r_0, r_5, V, R$. Then we replace the obtained values for $e^\alpha, \, e^\varsigma$ in the Hawking temperature in (\ref{th2}), obtaining
$$T_H=\frac{r_0^2 V \sqrt{R \left(\sqrt{g_s^2 Q_1^2+r_0^4 V^2}+g_s Q_1\right) \left(\sqrt{g_s^2 n^2+R^2 r_0^4 V^2}+g_s n\right)}}{2 \pi r_5 \left(\sqrt{g_s^2 Q_1^2+r_0^4 V^2}+g_s Q_1+r_0^2 V\right) \left(\sqrt{g_s^2 n^2+R^2 r_0^4 V^2}+g_s n+R r_0^2 V\right)}.$$

We could solve this expression for $r_0,$ as we did previously, but the solution, although possible, is notoriously lengthy. For our purposes it is enough to expand this expression for $T_H$ around $r_0=0$, obtaining
\be
T_H= \sqrt{\frac{R}{n Q_1}} \frac{V}{4 \pi g_s r_5} r_0^2 \left(1 - \frac{V}{2 g_s} \left(\frac{1}{Q_1}+ \frac{R}{n}\right) r_0^2 \right) + {\mathcal{O}} \left(r_0^5\right).
\ee
We see that the extremal limit corresponds geometrically to $r_0 \rightarrow 0,$ as expected, but in such a way that $\frac{r_0^2}{T_H}$ remains finite. As in the previously studied case, we conclude that the low frequency absorption cross section, given by (\ref{seccaowl2}), is the quotient of a term that goes to 0 in the extremal limit and the black hole temperature; and that quotient has a smooth extremal limit.

\subsection{Four dimensional black holes with two large charges}
\label{4lc}
\indent

Finally we consider a four dimensional dyonic four-charged black hole solution \cite{Cvetic:1995uj}, which also has a possible interpretation in terms of M-branes, corresponding to an eleven-dimensional configuration consisting of two 2-branes intersecting at a point, and two 5-branes intersecting at a 3-brane, with each of the 2-branes intersecting with each of the 5-branes at a string, and after compactification on $T^7$ \cite{Cvetic:1996gq}.
The metric given by
\be
ds_4^2 = -h^{-1/2}(r) f(r) d t^2 + h^{1/2}(r) \left( \frac{dr^2}{f(r)} + r^2 d \Omega_{2}^2 \right), \label{g4}
\ee
with the thermal and harmonic functions defined in $d=4$ as
\bea
f(r) &=& 1 - \frac{r_0}{r}, \\
h(r) &=& h_1(r) h_2(r) h_3(r) h_n(r), \nonumber \\
h_n(r) &=& 1 + \frac{r_n}{r}, \, h_i(r) = 1 + \frac{r_i}{r}, \, i=1, 2, 3.
\eea
The structure is very similar to the five-dimensional solution (\ref{g5}), necessarily adapted (like the functions $f(r), \, h_i(r),\, h_n(r)$) to $d=4$ and to four charges.

We are considering the range of parameters
\be
r_0, r_1, r_n \ll r_2, r_3,
\ee
Physically, this range of parameters means the charges associated to $r_2, r_3$ are much larger than those associated to $r_1, r_n,$ as before. Also as before, it is useful to introduce hyperbolic angles $\alpha, \varsigma$ relating the small radii $r_1, r_n,$ and the extremality parameter $r_0,$ but this time, in $d=4$, in the form
\be
r_1 \equiv r_0 \sinh^2\alpha, \, r_n \equiv r_0 \sinh^2 \varsigma.
\ee
In this range of parameters, the Bekenstein-Hawking entropy coincides with the entropy of a gas of effective strings with central charge $c=6$ and tension $T_{\rm eff} =\frac{1}{8\pi r_2 r_3}.$ The mass levels are given by
$$m^2= \left(2 \pi n_w T_{\rm eff}+n_p \right)^2 +8\pi T_{\rm eff} N_R=
\left(2 \pi n_w T_{\rm eff}-n_p \right)^2 +8\pi T_{\rm eff} N_L,$$
with (here $\kappa_4^2= 8 \pi G_4$ is Newton's constant in four dimensions)
\bea
m &=& \frac{\pi r_0}{\kappa_4^2} (\cosh(2\alpha)+\cosh(2\varsigma)), \\
n_w &=& 4 r_2 r_3 \frac{\pi r_0}{\kappa_4^2} \sinh(2\alpha), \label{nw4} \\
n_p &=& \frac{\pi r_0}{\kappa_4^2} \sinh(2\varsigma). \label{np4}
\eea
In this gas of effective strings once more we have left and right oscillators in a thermal form. The corresponding left and right temperatures, and from (\ref{thtltr}) the Hawking temperature, are in this case given respectively by
\bea
T_L &=& \frac{1}{4\pi \sqrt{r_2r_3} \cosh(\alpha - \varsigma)}, \, T_R= \frac{1}{4\pi \sqrt{r_2r_3} \cosh(\alpha+ \varsigma)}, \\
T_H &=& \frac{1}{4\pi \sqrt{r_2r_3} \cosh \alpha \cosh \varsigma}. \label{tltrth3}
\eea

Once again in this range of parameters it is possible to compute the full frequency-dependent scalar absorption cross section, which is given by \cite{Klebanov:1997cx}
\be
\sigma_{abs}(\omega)=2\pi\sqrt{r_2r_3} r_0 (\cosh 2\alpha +\cosh 2\varsigma) \omega \frac{e^{\frac{\omega}{T_H }} -1}{ \left(e^{\frac{\omega}{2T_L }} -1\right)\left(e^{\frac{\omega}{2T_R }} -1\right)}. \label{seccaow3}
\ee
Its low frequency limit is given by
\be
\sigma = \frac{r_0}{T_H}. \label{seccaowl3}
\ee

In order to study the extremal limit, we proceed as in the previous section: first we invert (\ref{nw4}) (respectively (\ref{np4})) in order to express $e^\alpha$ (respectively $e^\varsigma$) in terms of $n_w$ (respectively $n_p$), $r_0, r_2, r_3$. Then we replace the obtained values for $e^\alpha, \, e^\varsigma$ in the Hawking temperature in (\ref{tltrth3}), obtaining

$$T_H=\frac{r_0 \sqrt{\left(\sqrt{\kappa^8 n_p^2+4 \pi ^2 r_0^2}+\kappa^4 n_p\right) \left(\sqrt{\kappa^8 n_w^2+64 \pi ^2 r_0^2 r_2^2 r_3^2}+\kappa^4 n_w\right)}}{\left(\sqrt{\kappa^8 n_p^2+4 \pi ^2 r_0^2}+\kappa^4 n_p+2 \pi  r_0\right) \left(\sqrt{\kappa^8 n_w^2+64 \pi ^2 r_0^2 r_2^2 r_3^2}+\kappa^4 n_w+8 \pi  r_0 r_2 r_3\right)}.$$
Expanding this expression for $T_H$ around $r_0=0$, we obtain
\be
T_H=\frac{r_0}{2 \kappa^4 \sqrt{n_p n_w}} \left(1-\frac{\left(n_w + 4 r_2 r_3 n_p\right) }{\kappa^4 n_p n_w} \pi r_0 \right) + {\mathcal{O}} \left(r_0^3\right).
\ee
The previous discussion about the extremal limit can also be taken here, with similar conclusions. The extremal limit corresponds geometrically to $r_0 \rightarrow 0,$ as expected, but in such a way that $\frac{r_0}{T_H}$ remains finite. The low frequency absorption cross section in (\ref{seccaowl3}) is given by the quotient of a term that goes to 0 in the extremal limit and the black hole temperature; that quotient has a smooth extremal limit.

\subsection{General discussion}
\indent

For all the black holes we have considered in this section, in a string theory context, the low frequency absorption cross section has a form which is similar to (\ref{seccaocl}), with an explicit dependence of the temperature of the form $1/T_H.$ (Because the remaining terms in the expressions for all the cross sections we have considered depend implicitly on the temperature, we cannot say there is an inverse proportionality to $T_H.$)

This explicit dependence had been previously suggested in \cite{Halyo:1996xe}, where it is assumed that fundamental string states are in one to one correspondence with black hole states: the states of a free string evolve into black holes as the string coupling $g_s$ is turned on and increased; conversely, every black hole evolved from a state of a single free string. Based on these assumptions, the microscopic and macroscopic entropies can be related and shown to be equal to the black hole horizon area, for which the following formula was proposed:
\be
A_H= \lim_{{\omega}\rightarrow 0} \frac{2 \pi^2 P(\omega)}{\omega^2 T_H}, \label{hh}
\ee
with $P(\omega)$ being the low energy power spectrum of the Hawking radiation emitted by the black hole of mass $M$. Under the same assumptions, also in \cite{Halyo:1996xe} an expression for this power spectrum was obtained:
\be
P(\omega)=\frac{32 G}{M} T_L T_R \omega^2. \label{pw}
\ee
Replacing (\ref{pw}) in (\ref{hh}), and recalling (\ref{seccaocl}), one finds for the low frequency absorption cross section
\be
\sigma= \frac{64 \pi^2 G}{M} \frac{T_L T_R}{T_H},
\ee
a result which exhibits the expected explicit temperature dependence.

That dependence results from the fact that all cross sections we have seen were of the form
\be
\sigma_{abs}(\omega)=g_{eff} \, \omega \, \frac{\rho \left(\frac{\omega}{2 T_L} \right) \rho \left(\frac{\omega}{2 T_R} \right)}{\rho \left(\frac{\omega}{T_H} \right)}, \label{seccaowg}
\ee
where $g_{eff}$ is a (charge-dependent but frequency-independent) effective coupling of left and right moving oscillations of energies $\omega/2$ to an outgoing scalar of energy $\omega$, and the thermal factor $\rho(\omega/T)$ is given by
\be
\rho \left(\frac{\omega}{T} \right)\equiv \frac{1}{e^{ \frac{\omega}{ T}}  -1}.
\ee
Indeed in \cite{Das:1996wn,Maldacena:1996ix} it was shown that the total rate of emission of low energy scalar quanta by the D1-D5 configuration we considered is generically given by $$\Gamma(\omega)= g_{eff} \, \omega \, \rho\left(\frac{\omega}{2 T_L} \right) \, \rho \left(\frac{\omega}{2 T_R} \right) \frac{d^d k}{(2\pi)^d}.$$
Together with the detailed balance condition
$$\Gamma(\omega)= \sigma_{abs}(\omega) \rho \left(\frac{\omega}{T_H} \right) \frac{d^d k}{(2\pi)^d}$$
we easily obtain (\ref{seccaowg}), which is the generic form of the absorption cross section for the kind of stringy black holes we considered. The low frequency limit of (\ref{seccaowg}) is given by $\sigma = 4 g_{eff} \frac{T_L T_R}{T_H},$ which indeed has the aforementioned explicit dependence on the Hawking temperature.

\subsection{Inclusion of $\a$ corrections}
\indent

All black hole solutions we have considered in this section have their origin in string theory; yet, they do not include any $\a$ corrections. On the other hand, the analysis we have made in section \ref{smcsss} is valid for $\a$--corrected black holes. It is interesting to compare the results. In both cases, and from different and independent arguments, we are led to expect an explicit temperature dependence of the absorption cross section of the form $1/T_H$.

The $\a$ corrections to the solutions we have considered in this section are not known. Still, from the general result obtained in section \ref{smcsss}, even not knowing such corrections we expect the same explicit temperature dependence of the $\a$-corrected absorption cross section.

When deriving $\a$ corrections to black hole solutions, it is possible to choose the radial coordinate in such a way that the black hole horizon radius parameter $R_H$ does not have any $\a$ corrections. Exactly the same choice can be made, when determining the $\a$ corrections to each of the solutions considered in this section, to the extremality parameter $r_0$. Having guaranteed that choice has been made in each solution, from the result (\ref{seccaoto}) we conclude that the low frequency limits of the cross sections we have obtained in subsections \ref{2lc} and \ref{1lc} (given in both cases by the same formula (\ref{seccaowl2})) and in subsection \ref{4lc} (given by (\ref{seccaowl3})) are also valid in the presence of $\a$ corrections, if they keep being written on those forms (in terms of $r_0$), exactly in the same as (\ref{seccaoto}) is written in terms of $A_H$. If in any of those formulas one replaces $r_0$ by the black hole charges, explicit $\a$ corrections should appear.

\section{Conclusions}
\indent

In this article, we have shown that the low frequency absorption cross section for charged spherically symmetric $d$--dimensional black holes has an explicit dependence on the black hole temperature of the form $1/T_H$. Such explicit dependence appears as naturally induced (and required) by $\a$ corrections to the metric.

The validity of this assertion is also supported by considering intrinsically stringy solutions already in their classical, without $\a$ corrections limit: we have seen that, assuming fundamental string states to be in one to one correspondence with black hole states, at the classical level one naturally obtains a low frequency cross section with the same explicit dependence on the black hole temperature of the form $1/T_H.$ We verified this property for different regimes of the D1-D5 system in $d=5$ and a four dimensional dyonic four-charged black hole. From our result concerning $\a$-corrected black holes, we are led to the conclusion that the expressions for the low frequency cross section obtained from these solutions should also be valid in the presence of $\a$ corrections, even if we do not know such corrections to these solutions.

This way equation (\ref{seccaoto}), which seemed just a formula to encode the leading order $\a$ corrections to the low frequency absorption cross section, may actually be the most natural way to write this quantity in the context of string theory, since it is valid classically and with leading $\a$ corrections. For this to be true, one would have to show that equation (\ref{seccaoto}) remains valid in the presence of the full series of $\a$ corrections, and not just at leading order. Nonetheless, the matching we found up to leading order $\a$ corrections is remarkable.

For the solutions we considered, we have also evaluated the low frequency absorption cross section near the extremal limit. In each case we checked that, despite the $1/T_H$ dependence on the temperature, the cross section is well defined, giving results which in each case are equal to the horizon area, as it is expected for a classical solution. Taking this quantity in the extremal limit results in a smooth behavior. But for extremal black holes the string $\a$ corrections cannot be induced by the temperature; more generally, since nothing in that case can depend on the temperature, a formula like (\ref{seccaoto}) cannot be valid for extremal black holes and must be replaced. It remains an open issue to find a formula for the $\a$-corrected absorption cross section for extremal black holes. We address such issue in a forthcoming work \cite{Moura:2014lxa}.

\section*{Acknowledgments}
This work has been supported by FEDER funds through \emph{Programa Operacional Fatores de Competitividade -- COMPETE} and by Funda\c c\~ao para a Ci\^encia e a Tecnologia (FCT) through projects Est-C/MAT/UI0013/2011 and CERN/FP/123609/2011.


\bibliographystyle{plain}

\end{document}